\documentclass{article}

\usepackage[english]{babel}

\usepackage[letterpaper,top=2cm,bottom=2cm,left=3cm,right=3cm,marginparwidth=1.75cm]{geometry}

\usepackage{endnotes}
\usepackage{amssymb,amsmath,amsfonts,txfonts,mathdots}
\usepackage{textcomp}
\usepackage{braket}
\usepackage{soul}
\usepackage{graphicx}
\usepackage{authblk}
\usepackage[colorlinks=true, allcolors=blue]{hyperref}
\begin{document}
\title{Revisiting the Lindemann’s criterion from a minimal viscosity perspective}

\author[1]{Pablo G. Tello}
\affil[1]{CERN, Geneva, Switzerland}

\author[2,3,4]{Sauro Succi}
\affil[2]{Istituto per le Applicazioni del Calcolo "M. Picone", CNR, I-00185, Rome, Italy}
\affil[3]{Istituto Italiano di Tecnologia, 00161, Rome, Italy}
\affil[4]{Physics Department, Harvard University, 17 Oxford St, Cambridge, MA 02138, United States}

\maketitle

\begin{abstract}
A revisiting of the Lindemann criterion under the recently established minimal 
viscosity formulation is proposed which uncovers intriguing insights into the melting process. 
The approach suggests that melting involves competition between the electron rest 
energy and the thermal one caused by temperature increase when considering a 
defect-free crystal structure. 
However, when accounting for the role of vacancies in the melting process, the analysis 
suggests that the key competing factors are the thermal energy and the one needed for 
vacancy migration. The estimated value of the Lindemann constant, in this case, is close to 
the reported values and with recent simulations studying the correlation between 
Born and Lindemann melting criteria.
\end{abstract}

\section{Introduction}

Melting is a profoundly intricate phenomenon that continues to elude a
comprehensive theoretical understanding \cite{dash1999history,
phillpot1989crystals,kuhlmann1965theory,
stishov1975thermodynamics}. 
Over the course of
several decades, three primary explanatory frameworks have been put
forth. The first, attributed to Lindemann, posits that melting is
instigated by the onset of an instability triggered when the atomic
displacements during thermal vibrations surpass a critical threshold
known as the "Lindemann criterion'' \cite{lindemann1910calculation}. The second, credited to
Born, contends that melting results from a rigidity catastrophe stemming
from the vanishing of the elastic shear modulus; essentially, when the
crystal can no longer maintain sufficient rigidity to withstand external
forces, it transforms into a liquid state referred to as the "Born
instability" \cite{born1939thermodynamics}. Finally, in the interpretations proposed by Cahn
and others, the spontaneous generation of intrinsic lattice defects,
such as vacancies and arrays of intrinsic dislocations in proximity to
the melting point, is believed to be the driving force behind the
disintegration of long-range crystalline order \cite{cahn1986nucleation}.

Recent simulations reveal a compelling correlation between the Lindemann
and Born criteria in the onset of melting, suggesting that local lattice
instabilities, governed by both criteria, play a pivotal role
\cite{jin2001melting,
cahn2001materials,
zhou2005bridging}. It remains remarkable though that, notwithstanding the
intricate nature of the melting process, the Lindemann criterion
persists as a robust phenomenological approximation that displays only a
modest variation of approximately 10\% within a given crystal structure
type, spanning an overall range from 0.068 to 0.114 \cite{dash1999history,
phillpot1989crystals,
kuhlmann1965theory,
stishov1975thermodynamics}.

As previously mentioned, Lindemann postulated that during the melting
phase, the characteristic vibrational displacements specific to a
particular crystalline class should maintain a consistent fraction of
the lattice spacing. He proposed that this ratio should approximate half
of the lattice spacing, implying direct collisions among the
lattice' s constituent atoms, ultimately leading to the
breakdown of its ordered structure. It is worth noting though that this
ratio was further refined by Gilvarry to be approximately one-tenth of
the lattice spacing which will play a significant role in the coming
sections \cite{gilvarry1956lindemann}.

\section{Lindemann criterion and minimal viscosity}
The objective of this section is to examine the Lindemann criterion
through the lens of minimal viscosity \cite{trachenko2020minimal,
trachenko2021quantum}. The rationale behind
this will become evident as we delve into this section and the
subsequent ones. For now, let us consider that viscosity, in its general
sense, signifies the resistance to deformation at a specific rate. As
per the Lindemann criterion, the initiation of melting occurs when atoms
deviate from their equilibrium positions. Consequently, the concept of
minimal viscosity may offer insights into understanding the various
factors in play during the melting process. To put it simply, our goal
is to leverage certain assumptions applied to the simplicity of the
Lindemann criterion, in conjunction with the concept of minimal
viscosity, to elucidate the factors qualitatively and approximately in
competition during the melting process. Let us start by briefly
describing the reasoning underlying Lindemann criterion. The idea behind
is the observation that the average amplitude of thermal vibrations
increases as temperature does. Therefore, melting initiates when the
amplitude of vibration becomes large enough for adjacent atoms to partly
occupy the same space. The Lindemann criterion states that melting is
expected when the root mean square vibration amplitude exceeds a
threshold value. Assuming that all atoms in a crystal vibrate with the
same frequency \emph{f}, the average thermal energy can be estimated
using the equipartition theorem as:

\begin{equation} \label{GrindEQ__1_} 
E=4{\pi }^2mf^2u^2=k_BT 
\end{equation} 
Where \textit{$m$} is the atomic mass, \textit{$\nu$} is the frequency, \textit{u} is the average vibration amplitude, \textit{k${}_{B}$} is the Boltzmann constant, and \textit{T} is the absolute temperature. If the threshold value of \textit{u${}^{2}$ }is \textit{c${}^{2}$a${}^{2}$}, where \textit{c} is the Lindemann constant and \textit{a} is the atomic spacing, then the melting temperature is estimated as:
\begin{equation} \label{GrindEQ__2_} 
T_m=\frac{4{\pi }^2mf^2u^2c^2}{k_B} 
\end{equation} 
For our purposes, we formulate Lindemann criterion \eqref{GrindEQ__2_} in its usual way considering the mean square displacement in one spatial direction, \textit{$<$u${}^{2}$$>$${}_{AV}$}:
\begin{equation} \label{GrindEQ__3_} 
{\left\langle u^2\right\rangle }_{AV}=\frac{3k_BT_m}{4{\pi }^2m{\omega }^2_D} 
\end{equation} 
Where \textit{$\omega$${}_{D}$} is the Debye frequency (\textit{$\omega$${}_{D}$=k${}_{B}$$\theta$${}_{D}$/h} with \textit{$\theta$${}_{D}$} being the Debye temperature\textit{)}. Instead of \textit{u${}^{2}$$\sim$ c${}^{2}$a${}^{2}$}, we will reformulate Lindemann assumption as \textit{$<$u${}^{2}$$>$${}_{AV}$ $\sim$ $\delta$${}^{2}$a${}_{B}$${}^{2}$} by considering Bohr's radius, \textit{a${}_{B}$}. This essentially involves redefining the average vibration amplitude by considering a fixed minimum length, such as Bohr's radius. It is important to note that in crystalline solids, the interatomic distance is typically on the order of angstroms, which is roughly one order of magnitude greater than Bohr's radius. Therefore, our approximation is well-founded within the qualitative objectives and scope of this paper. Furthermore, Gilvarry's refinement of the Lindemann criteria to one-tenth of the lattice spacing lends it further credibility as it aligns with the magnitude of Bohr's radius (\textit{c$\sim$10${}^{-10}$} m vs \textit{a${}_{B}$$\sim$10${}^{-11}$ m}). Thus, the previous expression becomes:
\begin{equation} \label{GrindEQ__4_} 
{\delta }^2a^2_B\sim \frac{k_BT_m}{m{\omega }^2_D} 
\end{equation}

\noindent 

\noindent Let us now consider the dynamic viscosity derived by Andrade for a monomolecular liquid at the melting point, which is \cite{andrade1952viscosity}:
\begin{equation} \label{GrindEQ__5_} 
{\eta }_M\sim \frac{m}{\sigma }f_M 
\end{equation} 
Where \textit{m} is the mass of a molecule, \textit{$\sigma$} the average distance between the centers of molecules, and \textit{f${}_{M}$}${}_{\ }$the vibration frequency at melting. The relationship between the dynamic viscosity, \textit{$\eta$${}_{M}$ , }and the kinematic one\textit{, $\upsilon$${}_{M}$, }is given by\textit{ $\eta$${}_{M}$ = $\upsilon$${}_{M\ }$$\rho$}, where \textit{$\rho$} is density (\textit{$\rho$$\sim$m/$\sigma$${}^{3}$}).  Therefore, expression \eqref{GrindEQ__5_} becomes:
\begin{equation} \label{GrindEQ__6_} 
{\upsilon }_M\sim f_M{\sigma }^2 
\end{equation} 
We now approximate \textit{$\sigma$ to the }Bohr radius\textit{, a${}_{B}$ }and the vibration frequency at melting\textit{, f${}_{M}$, to the }Debye frequency\textit{, }w\textit{${}_{D}$. }Therefore, expression \eqref{GrindEQ__6_} becomes ${\nu }_M={\omega }_Da^2_B$.

\noindent Following the considerations reported by Trachenko et al., the Debye frequency can, in turn, be related to the characteristic bonding strength set by the Rydberg energy \cite{trachenko2020minimal,trachenko2021quantum}. This leads, in our case to  
\begin{equation} \label{GrindEQ__7_} 
{\upsilon }_M\sim \frac{\hslash }{{\left(m_em\right)}^{{1}/{2}}} 
\end{equation} 
which corresponds to the expression for the minimum kinematic viscosity derived by Trachenko et al. \cite{trachenko2020minimal,trachenko2021quantum}.  Considering now \eqref{GrindEQ__6_} and \eqref{GrindEQ__7_} we have \textit{$\omega$${}_{D}$$\sim$$\textrm{h}$ /a${}_{B}$${}^{2}$(m${}_{e}$m)${}^{1/2}$}. By substituting this into expression \eqref{GrindEQ__4_}, we obtain \textit{$\delta$${}^{2}$$\sim$m${}_{e}$a${}_{B}$${}^{2}$(k${}_{B}$T${}_{m}$/h${}^{2}$)} and by recalling that the Bohr radius is approximately \textit{a${}_{B}$$\sim$h/$\alpha$m${}_{e}$c}, being \textit{$\alpha$} the fine structure constant, we finally obtain:
\begin{equation} \label{GrindEQ__8_} 
{\delta }^2\sim \frac{{k_BT}_m}{{\alpha }^2m_ec^2} 
\end{equation} 
By considering \textit{T${}_{m}$=10${}^{3}$ K}, therefore \textit{k${}_{B}$T${}_{m}$ $\sim$0.1 eV}, and \textit{m${}_{e}$c${}^{2}$ $\sim$ 0.5 MeV}, and 1/\textit{$\alpha$${}^{2}$$\sim$10${}^{4}$}, a value of \textit{$\delta$$\sim$0.04 }which is lower than the Lindemann constant value range mentioned previously. This could be attributed to the heuristic approximation taken. The main merit of expression \eqref{GrindEQ__8_} though is manifesting that melting appears to be ultimately a competition between the electron rest energy against the thermal one given by the increase of temperature when considering a defect-free crystalline structure. In the next section we will address the role of the vacancies during the melting process and as it will be shown, the values of \textit{$\delta$} will get into a closer agreement to the ones reported for the Lindemann constant. 

\section{Role of the vacancies}

\noindent As long established theoretically and experimentally, there is a correlation 
between the formation and migration of vacancies in metals and their melting points. 
Evidence shows that the onset of melting occurs when the concentration of vacancies within the solid metal reaches a critical threshold. Therefore, melting can be described as the generation of additional vacancies and its migration, driven by the release of latent heat during fusion \cite{gorecki1977comments,tiwari1975correlation,mei2007melting}. Within this context, let us now consider, along the same reasoning line as in the previous section, the role that vacancies might play and the onset of melting. As already reported elsewhere, the energy of vacancy migration \textit{E${}_{m}$} could be written as:
\begin{equation} \label{GrindEQ__9_} 
E_m\sim {\left(\frac{k_B}{\hslash }\right)}^2m{\theta }^2_Da^2 
\end{equation} 
Where \textit{m} is the atomic mass, \textit{$\theta$${}_{D}$} the Debye temperature and \textit{a} the interatomic spacing \cite{couchman1976relation,glyde1967relation}. As in the previous section we will assume \textit{a$\sim$a${}_{B}$}. Taking into account the relationship between the Debye temperature and frequency, \textit{$\theta$${}_{D}$=h$\omega$${}_{D}$/k${}_{B}$}, we obtain \textit{E${}_{m}$$\sim$m$\omega$${}_{D}$${}^{2}$a${}_{B}$${}^{2}$}, hence\textit{ a${}_{B}$${}^{2}$$\sim$E${}_{m}$/m$\omega$${}_{D}$${}^{2}$}. By considering expression \eqref{GrindEQ__4_}, we finally arrive to:
\begin{equation} \label{GrindEQ__10_} 
{\delta }^2\sim \frac{{k_BT}_m}{E_m} 
\end{equation} 
The first observation to be noticed by comparing the obtained expression with \eqref{GrindEQ__8_} 
is that now the competing factors become the thermal energy vs. the energy required 
for the migration of vacancies. 
Expressed in other words, the thermal energy vs the one needed for an atom, originally in its ordered 
place within the crystalline structure, to jump towards an empty lattice place. 
It should be noticed that the range of vacancy migration energies could be rather 
wide (e.g. 0.08 eV for fcc K to 2.46 eV for hcp Ru) \cite{shang2016comprehensive}. 
Roughly considering the average between these two values (1.27 eV) and using it in 
equation \eqref{GrindEQ__10_}, a value of \textit{$\delta$ $\sim$ 0.28 }is obtained, which is 
in better agreement with the one mentioned in the introduction.
 Remarkably, it is noticeable the closeness of this value to the one obtained by Jin et al. in their 
simulations \cite{jin2001melting}. 
They named as ``Lindemann atoms'' those ones for which their fractional root-mean-square 
displacement exceeded the critical value \textit{$\delta$$\sim$0.22}.
 As they showed, the amount and location of these Lindemann atoms increased rapidly and 
drastically towards approaching the melting temperature and formed clusters of increasing size as the crystal heated up. 
Moreover, Jin et al. calculated the Born elastic moduli for the Lindemann atoms, finding that the 
average value of their elastic shear modulus was much closer to zero for the totality of these atoms 
than for the whole crystal.  

It is also of interest to associate the expression (14) to an Arrhenius-like probability 
$p(\delta) \propto e^{-1/\delta^2}$ (normalized to a finite interval, say $[0,1]$,
that a fluctuation of relative amplitude \textit{$\delta$} manages to overcome the vacancy 
migration barrier, thereby contributing a ``melting event''. 

Such probability is heavily (non-analytically) suppressed up to 
values of \textit{$\delta$$\sim$0.3}; 
for instance, \textit{p(0.1)$\sim$10${}^{-43}$}, \textit{p(0.2)$\sim$10${}^{-11}$}, 
\textit{p(0.3)$\sim$10${}^{-5}$}, \textit{p(0.5) $\sim$10${}^{-2}$}, 
\textit{p(1)$\sim$3$\cdot$10${}^{-2}$}. 

Based on this estimate,\textit{ $\delta$$\sim$0.28} lies just at the lower end of 
the transition region where the probability \textit{p($\delta$)} starts to build up 
substantial values, whereas the value \textit{$\delta$$\sim$0.1} sits down deep into the ``un-melting'' region. 
This further highlights the importance of taking into account realistic values of 
the vacancy migration energy (see Figure 1). 

\begin{figure}
\includegraphics[scale=0.5]{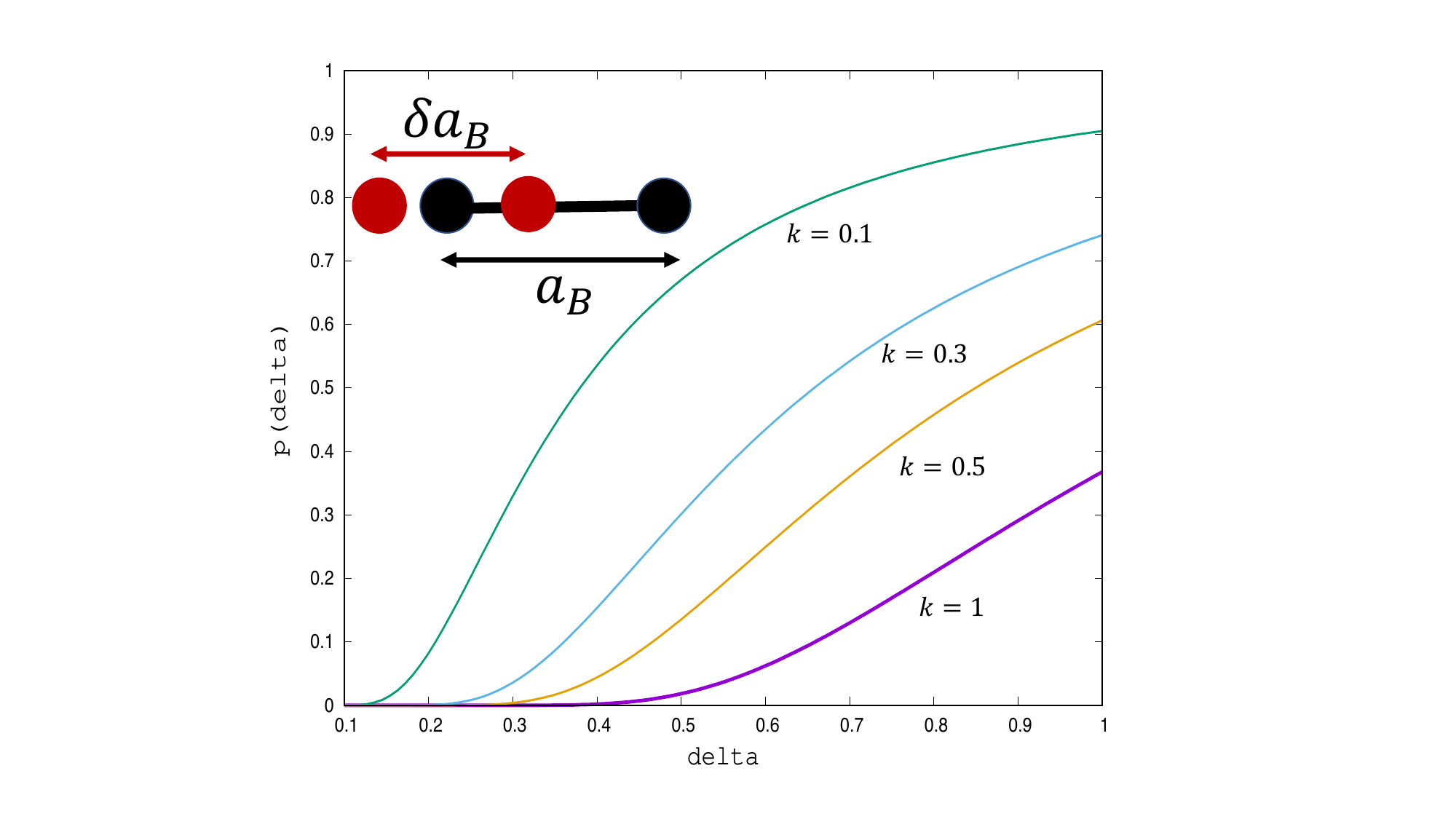}
\caption{The probability $p(\delta;k) = e^{-k/\delta^2}$ for $k=0.1,0.3.0.5,1$.
Interestingly, for $k=1$, the probability starts to take sizeable 
values for $\delta > 0.5$, which is the natural threshold for melting.
However, the non-analytic dependence at $\delta \to 0$ makes $p(\delta)$ 
very sensitive to the value of the parameter $k$, which highlights the importance
of taking into account realistic values of the migration energy. 
}
\end{figure}

In closing, we note that such probabilistic description of atomic displacements may prove 
useful for the development of coarse-grained computational models of transport 
phenomena in the presence of melting and solidification 
\cite{hu1996mathematical,rasin2005phase} and in porous media 
as well \cite{bear2018modeling,cali1992diffusion}.
For instance one could track the spacetime dependence of $\delta$ as a function of the
local temperature and use $p(\delta)$ as a probabilistic threshold to decide about 
solidification/melting events in kinetic Monte Carlo simulations or lattice Boltzmann simulations.

\section{Conclusions}
We have revisited the Lindemann criterion, under the light of the general notion of minimal viscosity. 
This novel perspective reveals interesting aspects of the melting process. 
The first one is that melting appears to involve a fundamental competition between the electron 
rest energy against the thermal one, given by the increase of temperature when considering a 
defect-free crystalline structure. 
The obtained value for the Lindemann constant is lower than the range of the traditionally reported ones. 
We have also considered the role of the vacancies during the melting process. 
The results obtained indicate that the competing factors during melting are the thermal energy vs. 
the energy required for the migration of vacancies. 
The estimated value of the Lindemann constant, in this case, is closer to the reported ones, as well 
as to the result of recent simulations investigating the correlation between Lindemann and Born melting criteria.

\section*{Acknowledgments}
The authors would like to specially acknowledge the valuable input and critical reading 
from Prof. K. Trachenko from the School of Physical and Chemical Sciences, Queen Mary University 
of London, during the elaboration of this manuscript. 
SS wishes to acknowledge financial support from the ERC-PoC grant DROPTRACK (101081171).

\bibliographystyle{ieeetr}
\bibliography{sample}

\end{document}